\documentclass[preprint,authoryear]{elsarticle}
\usepackage{multirow}
\usepackage{color}
\usepackage{amsmath}
\usepackage{amssymb}
\usepackage{caption}
\usepackage{subcaption}
\usepackage{epstopdf}
\usepackage[colorinlistoftodos,textsize=small]{todonotes}
\usepackage[breaklinks]{hyperref}
\usepackage{tikz}
\usepackage{hhline}

\graphicspath{ {./figures/} }

\begin{document}

\begin{frontmatter}
\title{Corrigendum to \textquotedblleft Random grid-based visual secret sharing with abilities of OR and XOR decryptions\textquotedblright[Journal of Visual Communication and Image Representation. 24(2013) 48-62]}
\author{Saurabh Bora\corref{cor1}\fnref{fn1}}
\ead{saurabh.bora@iiitdmj.ac.in}
\cortext[cor1]{Corresponding author}
\fntext[fn1]{Student}
\author{Aparajita Ojha\fnref{fn2}}
\ead{aojha@iiitdmj.ac.in}
\fntext[fn2]{Professor}
\address{PDPM Indian Institute of Information Technology, Design and Manufacturing Jabalpur, India}

\begin{abstract}
  It has been observed that the contrast values for (2, 3) VSS scheme, (2, 4) VSS scheme, (3, 5) VSS scheme and (4, 5) VSS scheme claimed by \cite{Wu201348} are incorrect. Since the same values are cited and compared by many other researchers in their works, we have calculated and presented the correct values of contrast in this note.
\end{abstract}

\begin{keyword}
Visual Secret Sharing \sep Random Grid \sep Threshold \sep Visual Cryptography \sep OR \sep XOR \sep Contrast \sep Shares
\end{keyword}
\end{frontmatter}

\section{\label{sec:introduction}Introduction}
\cite{Wu201348} have recently introduced a $(k,n)$ VSS scheme which has capability of OR and XOR decryption both simultaneously. For large number of shares, the reconstructed secret image by OR operation has a very low contrast. However, if a light weight computational device is available, the secret image can be reconstructed using XOR operation which has higher contrast. The scheme works very well, but the value of contrast mentioned in the paper are erroneous due to possible oversight in computation by the authors.

The rest of the note is organized as follows. In Section 2 we have computed the correct contrast values for (2, 3) VSS scheme as an example and presented the corrected contrast values for all other schemes. Finally, we conclude in section 3.

\section{\label{sec:Contrast}Contrast Computation}
Contrast values as depicted in \cite{Wu201348}(Table 5, Table 6, Table 8, Table 9) are incorrect. In the following, we present the correct calculations and the final corrected values of contrast for (2, 3), (2, 4), (3, 5) and (4, 5) VSS schemes.\\
Definitions and formulae used in present note are the same as those introduced by \cite{Wu201348} and \cite{Chen20111197}. For various VSS schemes, we have computed the values of contrast for OR and XOR decryption both in Appendix using those formulae.\\ 
So, we calculate the value of contrast by OR decryption for (2, 3) VSS scheme. Consider a secret pixel s, n be total number of pixels generated corresponding to each secret pixel, t is the number of pixels to be stacked, r is the reconstructed secret pixel. As the secret can be reconstructed by stacking 2 or 3 shares, t can be 2 or 3.\\
case 1 : t = 2\\ 
Average light transmission when $s=0$,
\begin{align*}
T^{OR,2}(r[s=0]) &= \frac{1}{(n-k+1)} [T^{OR,2}_{(k,n)} (r[s=0]) + T^{OR,2}_{(k+1,n)} (r[s=0]) + ..\\ &\phantom{{}=\frac{1}{(n-k+1)} [T^{OR,2}_{(k,n)} (r[s=0])} ... + T^{OR,2}_{(n,n)} (r[s=0])]\\
&= \frac{1}{(3-2+1)} [T^{OR,2}_{(2,3)} (r[s=0]) + T^{OR,2}_{(3,3)} (r[s=0])]\\
&= \frac{1}{2} \times [ \frac {\binom 22}{\binom 32} \times (\frac{1}{2})^{2-1} +  (1-  \frac {\binom 22}{\binom 32})\times (\frac{1}{2})^{2} + (\frac{1}{2})^{2} ]\\
& = 7/24.
\end{align*}
Average light transmission when $s=1$, 
\begin{align*}
T^{OR,2}(r[s=1]) &= \frac{1}{(n-k+1)} [T^{OR,2}_{(k,n)} (r[s=1]) + T^{OR,2}_{(k+1,n)} (r[s=1]) + ..\\&\phantom{{}=\frac{1}{(n-k+1)} [T^{OR,2}_{(k,n)} (r[s=1])}... + T^{OR,2}_{(n,n)} (r[s=1])]\\
&= \frac{1}{(3-2+1)} [T^{OR,2}_{(2,3)} (r[s=1]) + T^{OR,2}_{(3,3)} (r[s=1])]\\
&= \frac{1}{2} \times [ (1-  \frac {\binom 22}{\binom 32})\times (\frac{1}{2})^{2} + (\frac{1}{2})^{2} ]\\
& = 5/24.
\end{align*}
Contrast is calculated as  
\begin{align*}
\alpha &= \frac{T^{OR,2}(r[s=0]) - T^{OR,2}(r[s=1])}{1 +T^{OR,2}(r[s=1])} = \frac{\frac{7}{24}-\frac{5}{24}}{1+\frac{5}{24}} =   \frac{2}{29}.
\end{align*}
case 2 : t = 3\\
Average light transmission when $s=0$,
\begin{align*}
T^{OR,3}(r[s=0]) &= \frac{1}{(n-k+1)} [T^{OR,3}_{(k,n)} (r[s=0]) + T^{OR,3}_{(k+1,n)} (r[s=0]) + ..\\&\phantom{{}=\frac{1}{(n-k+1)} [T^{OR,3}_{(k,n)} (r[s=0])}... + T^{OR,3}_{(n,n)} (r[s=0])]\\
&= \frac{1}{(3-2+1)} [T^{OR,3}_{(2,3)} (r[s=0]) + T^{OR,3}_{(3,3)} (r[s=0])]\\
&= \frac{1}{2} \times [ \frac {\binom 32}{\binom 32} \times (\frac{1}{2})^{3-1} +  (1-  \frac {\binom 32}{\binom 32})\times (\frac{1}{2})^{3} +\frac {\binom 33}{\binom 33} \times (\frac{1}{2})^{3-1} + \\&\phantom{{}=+\frac {\binom 33}{\binom 33} \times (\frac{1}{2})^{3-1}} (1-  \frac {\binom 33}{\binom 33})\times (\frac{1}{2})^{3} ]\\
&= 1/4.
\end{align*}
Average light transmission when $s=1$, 
\begin{align*}
T^{OR,3}(r[s=1]) &= \frac{1}{(n-k+1)} [T^{OR,3}_{(k,n)} (r[s=1]) + T^{OR,3}_{(k+1,n)} (r[s=1]) + ..\\&\phantom{{}=\frac{1}{(n-k+1)} [T^{OR,3}_{(k,n)} (r[s=1])}... + T^{OR,3}_{(n,n)} (r[s=1])]\\
&= \frac{1}{(3-2+1)} [T^{OR,3}_{(2,3)} (r[s=1]) + T^{OR,3}_{(3,3)} (r[s=1])]\\
&= \frac{1}{2} \times [ (1-  \frac {\binom 32}{\binom 32})\times (\frac{1}{2})^{3} + (1-  \frac {\binom 33}{\binom 33})\times (\frac{1}{2})^{3} ]\\
&= 0.
\end{align*}
Contrast is calculated as  
\begin{align*}
\alpha &= \frac{T^{OR,3}(r[s=0]) - T^{OR,3}(r[s=1])}{1 +T^{OR,3}(r[s=1])} = \frac{\frac{1}{4}-0}{1+0}  = \frac{1}{4}.
\end{align*}
Now, we calculate the value of contrast for (2, 3) VSS scheme by XORed decryption\\
case 1 : t = 2\\  
Average light transmission when $s=0$, 
\begin{align*}
T^{XOR,t}(r[s=0]) &= \frac{1}{(n-k+1)} [T^{XOR,t}_{(k,n)} (r[s=0]) + T^{XOR,t}_{(k+1,n)} (r[s=0]) + ..\\&\phantom{{}=\frac{1}{(n-k+1)} [T^{XOR,t}_{(k,n)} (r[s=0])}... + T^{XOR,t}_{(n,n)} (r[s=0])]\\
&=\frac{1}{(3-2+1)} [T^{XOR,2}_{(2,3)} (r[s=0]) + T^{XOR,2}_{(3,3)} (r[s=0])]\\
&=\frac{1}{(3-2+1)} [ \frac {1}{2} \times (1 + \frac {1}{\binom 32}) +\frac {1}{2}]\\
&= 7/12.
\end{align*}
Average light transmission when $s=1$, 
\begin{align*}
T^{XOR,t}(r[s=1]) &= \frac{1}{(n-k+1)} [T^{XOR,t}_{(k,n)} (r[s=1]) + T^{XOR,t}_{(k+1,n)} (r[s=1]) + ..\\&\phantom{{}=\frac{1}{(n-k+1)} [T^{XOR,t}_{(k,n)} (r[s=1])}... + T^{XOR,t}_{(n,n)} (r[s=1])]\\
&=\frac{1}{(3-2+1)} [T^{XOR,2}_{(2,3)} (r[s=1]) + T^{XOR,2}_{(3,3)} (r[s=1])]\\
&=\frac{1}{(3-2+1)} [ \frac {1}{2} \times (1 - \frac {1}{\binom 32}) +\frac {1}{2}]\\
&= 5/12.
\end{align*}
Contrast is computed as  
\begin{align*}
\alpha &= \frac{T^{XOR,2}(r[s=0]) - T^{XOR,2}(r[s=1])}{1 +T^{XOR,2}(r[s=1])} = \frac{\frac{7}{12}-\frac{5}{12}}{1+\frac{5}{12}} 
= \frac{2}{17}. 
\end{align*}
case 2 : t = 3\\ 
Average light transmission when $s=0$,
\begin{align*}
T^{XOR,t}(r[s=0]) &= \frac{1}{(n-k+1)} [T^{XOR,t}_{(k,n)} (r[s=0]) + T^{XOR,t}_{(k+1,n)} (r[s=0]) + ..\\&\phantom{{}=\frac{1}{(n-k+1)} [T^{XOR,t}_{(k,n)} (r[s=0])}... + T^{XOR,t}_{(n,n)} (r[s=0])]\\
&=\frac{1}{(3-2+1)} [T^{XOR,3}_{(2,3)} (r[s=0]) + T^{XOR,3}_{(3,3)} (r[s=0])]\\
&=\frac{1}{(3-2+1)} [\frac {1}{2} + \frac {1}{2} \times (1 + \frac {1}{\binom 33}) ]\\
&= 3/4.
\end{align*}
Average light transmission when $s=1$,
\begin{align*}
T^{XOR,t}(r[s=1]) &= \frac{1}{(n-k+1)} [T^{XOR,t}_{(k,n)} (r[s=1]) + T^{XOR,t}_{(k+1,n)} (r[s=1]) + ..\\&\phantom{{}=\frac{1}{(n-k+1)} [T^{XOR,t}_{(k,n)} (r[s=1])}... + T^{XOR,t}_{(n,n)} (r[s=1])]\\
&=\frac{1}{(3-2+1)} [T^{XOR,3}_{(2,3)} (r[s=1]) + T^{XOR,3}_{(3,3)} (r[s=1])]\\
&=\frac{1}{(3-2+1)} [\frac {1}{2} + \frac {1}{2} \times (1 - \frac {1}{\binom 33}) ]\\
&= 1/4.
\end{align*}
Contrast is calculated as  
\begin{align*}
\alpha &= \frac{T^{XOR,3}(r[s=0]) - T^{XOR,3}(r[s=1])}{1 +T^{XOR,3}(r[s=1])} = \frac{\frac{3}{4}-\frac{1}{4}}{1+\frac{1}{4}} =  \frac{2}{5}.
\end{align*}
Similarly, contrast values for (2, 4) VSS scheme, (3, 5) VSS scheme and (4, 5) VSS scheme are also computed in Appendix. Table \ref{table:finaltable} shows the comparison between reported contrast by \cite{Wu201348} and correctly computed contrast in this note.
\begin{table*}[!htb] 
\caption{Comparison between previously stated contrast and our calculated contrast}                  
\centering                                          
\begin{tabular}{|p{2 cm}|p{1 cm}||p{2 cm}|p{2 cm}||p{2 cm}|p{2 cm}|}                                       \hline                                               
VSS scheme & t & \multicolumn{2}{p{4 cm}||}{\cite{Wu201348} Contrast values} & \multicolumn{2}{p{4 cm}|}{Corrected Contrast values} \\[1ex]
\hhline{~~----}
 & &$\alpha_{OR}$ & $\alpha_{XOR}$ &$\alpha_{OR}$ & $\alpha_{XOR}$ \\ 
\hline                                           
\multirow{2}{*} {(2, 3)} & 2 & $\frac{1}{10}$ & $\frac{1}{6}$ & $\frac{2}{29}$ & $\frac{2}{17}$\\[1ex]
                         & 3 & $\frac{5}{17}$ & $\frac{2}{5}$ & $\frac{1}{4}$ & $\frac{2}{5}$ \\[1ex]
\hline 
\multirow{3}{*} {(2, 4)}& 2 & $\frac{1}{15}$  & $\frac{1}{9}$ & $\frac{2}{89}$ & $\frac{2}{53}$ \\[1 ex]
                        & 3 & $\frac{14}{107}$ & $\frac{2}{57}$ & $\frac{6}{105}$ & $\frac{2}{35}$ \\[1 ex]
                        & 4 & $\frac{11}{49}$ & $\frac{3}{8}$ & $\frac{1}{8}$ & $\frac{1}{4}$\\[1 ex]
\hline  
\multirow{3}{*} {(3, 5)}& 3 & $\frac{2}{269}$  & $\frac{2}{89}$ & $\frac{2}{269}$  & $\frac{2}{89}$  \\[1 ex]
                        & 4 & $\frac{3}{126} = \frac{1}{42}$ & $\frac{1}{27}$ & $\frac{1}{42}$ &  $\frac{1}{22}$  \\[1 ex]
                        & 5 & $\frac{1}{16}$ & $\frac{1}{4}$ &$\frac{1}{16}$ & $\frac{1}{4}$ \\[1 ex]
\hline 
\multirow{3}{*} {(4, 5)}& 4 & $\frac{2}{169}$  & $\frac{1}{29}$& $\frac{2}{169}$ & $\frac{2}{29}$  \\[1 ex]
                        & 5 & $\frac{1}{16}$ & $\frac{2}{5}$& $\frac{1}{16}$ & $\frac{2}{5}$ \\[1 ex]
\hline                  
\end{tabular} 
\label{table:finaltable}                                                      
\end{table*}

\section{\label{sec:conclusion}Conclusion}
In the corrigendum, the computation errors in \cite{Wu201348} are corrected.

\appendix
\section{\label{sec:appendix}Appendix}
Let $I$ be the secret image and $I'$ be its reconstructed image by stacking $t$ shares i.e., $I'_{i_1\otimes i_2...\otimes i_t}$ = $R_1\otimes...R_t$. The contrast of $I'$ w.r.t $I$ is defined by \cite{Wu201348}
\begin{align*}
\alpha &= \frac{T(I'_{i_1\otimes i_2...\otimes i_t}[I(0)]) - T(I'_{i_1\otimes i_2...\otimes i_t}[I(1)])}{1 + T(I'_{i_1\otimes i_2...\otimes i_t}[I(1)]}
\end{align*}
where $I(0)$ and $I(1)$ denote the region consisting of white(transparent) pixels and black(opaque) pixels respectively in $I$. Further, $I'[I(0)]$ and $I'[I(1)]$ is the area of all white and black pixels in the reconstructed image $I'$ corresponding to the regions $I(0)$ and $I(1)$.\\
In \cite{Wu201348} according to Lemma 1 the average light transmission of stacked result when $s = j$, where j = 0 or j = 1  is given as 
\begin{align*}
T^{OR,t}(r[s=j]) &=\frac{1}{(n-k+1)} [T^{OR,t}_{(k,n)} (r[s=j]) + T^{OR,t}_{(k+1,n)} (r[s=j]) + ..\\&\phantom{{}=\frac{1}{(n-k+1)} [T^{OR,t}_{(k,n)} (r[s=j])}... + T^{OR,t}_{(n,n)} (r[s=j])]
\end{align*}
$T^{OR,t}_{(k,n)}$ is average light transmission of stacked results by t number of pixel values by \cite{Chen20111197} threshold RG based VSS scheme\\
\begin{gather*}
\begin{align*}
T^{OR,t}_{(k,n)}(r[s=0])=
\begin{cases}
 \frac {\binom tk}{\binom nk} \times (\frac{1}{2})^{t-1} +  (1-  \frac {\binom tk}{\binom nk})\times (\frac{1}{2})^{t} & \text{for } t\geq k  ...\text{from \cite{Chen20111197}}\\& \phantom{{} =} \text{(p.1202, Lemma 7)}\\
(\frac{1}{2})^{t} & \text{for } t<k  ...\text{from \cite{Chen20111197}}\\& \phantom{{} = } \text{(p.1200, Lemma 5)}\\
\end{cases}
\end{align*}
\end{gather*}
\begin{gather*}
\begin{align*}
T^{OR,t}_{(k,n)}(r[s=1])=
\begin{cases}
 (1-  \frac {\binom tk}{\binom nk})\times (\frac{1}{2})^{t} & \text{for } t\geq k  ...\text{from \cite{Chen20111197}}\\& \phantom{{} =} \text{(p.1202, Lemma 7)}\\
(\frac{1}{2})^{t} & \text{for } t<k  ...\text{from \cite{Chen20111197}}\\& \phantom{{} =} \text{(p.1200, Lemma 5)}\\
\end{cases}
\end{align*}
\end{gather*}
In Wu and Sun\cite{Wu201348} according to Lemma 6, the average light transmission of XORed result when $s = j$  is given as 
\begin{align*}
T^{XOR,t}(r[s=j]) &=\frac{1}{(n-k+1)} [T^{XOR,t}_{(k,n)} (r[s=j]) + T^{XOR,t}_{(k+1,n)} (r[s=j]) + ..\\& \phantom{{} =\frac{1}{(n-k+1)} [T^{XOR,t}_{(k,n)} (r[s=j])}... + T^{XOR,t}_{(n,n)} (r[s=j])]
\end{align*}\\
$T^{XOR,t}_{(k,n)}$ is average light transmission of XORed results by t number of pixel values by Chen and Tsao's\cite{Chen20111197} threshold RG based VSS scheme\\
\begin{gather*}
T^{XOR,t}_{(k,n)}(r[s=0])=
\begin{cases}
\frac {1}{2} \times (1 + \frac {1}{\binom nk}) & \text{for } t = k  ...\text{from \cite{Wu201348} (p.52, Lemma 4)}\\
(\frac{1}{2}) & \text{for } t > k  ...\text{from \cite{Wu201348} (p.52, Lemma 4)}\\
(\frac{1}{2}) & \text{for } t < k  ...\text{from \cite{Wu201348} (p.52, Lemma 3)}\\
\end{cases}
\end{gather*}
\begin{gather*}
T^{XOR,t}_{(k,n)}(r[s=1])=
\begin{cases}
 \frac {1}{2} \times (1 - \frac {1}{\binom nk}) & \text{for } t = k  ...\text{from \cite{Wu201348} (p.52, Lemma 4)}\\
(\frac{1}{2}) & \text{for } t > k  ...\text{from \cite{Wu201348} (p.52, Lemma 4)}\\
(\frac{1}{2}) & \text{for } t < k  ...\text{from \cite{Wu201348} (p.52, Lemma 3)}\\
\end{cases}
\end{gather*}
In \cite{Wu201348}(Table 6) contrast for (2,4) VSS scheme are depicted incorrectly. So, using the above stated formulae, we calculate the value of contrast by OR decryption.\\
case 1 : t = 2\\
Average light transmission when $s=0$,
\begin{align*}
T^{OR,2}(r[s=0]) &= \frac{1}{(n-k+1)} [T^{OR,2}_{(k,n)} (r[s=0]) + T^{OR,2}_{(k+1,n)} (r[s=0]) + ..\\&\phantom{{}=\frac{1}{(n-k+1)} [T^{OR,2}_{(k,n)} (r[s=0])}... + T^{OR,2}_{(n,n)} (r[s=0])]\\
&= \frac{1}{(4-2+1)} [T^{OR,2}_{(2,4)} (r[s=0]) + T^{OR,2}_{(3,4)} (r[s=0]) + \\&\phantom{{}= \frac{1}{(4-2+1)} [T^{OR,2}_{(2,4)} (r[s=0])}T^{OR,2}_{(4,4)} (r[s=0])]\\
&= \frac{1}{3} \times [ \frac {\binom 22}{\binom 42} \times (\frac{1}{2})^{2-1} +  (1-  \frac {\binom 22}{\binom 42})\times (\frac{1}{2})^{2} + (\frac{1}{2})^{2} + (\frac{1}{2})^{2}]\\
& = 19/72.
\end{align*}
Average light transmission when $s=1$, 
\begin{align*}
T^{OR,2}(r[s=1]) &= \frac{1}{(n-k+1)} [T^{OR,2}_{(k,n)} (r[s=1]) + T^{OR,2}_{(k+1,n)} (r[s=1]) + ..\\&\phantom{{}=\frac{1}{(n-k+1)} [T^{OR,2}_{(k,n)} (r[s=1])}... + T^{OR,2}_{(n,n)} (r[s=1])]\\
&= \frac{1}{(4-2+1)} [T^{OR,2}_{(2,4)} (r[s=1]) + T^{OR,2}_{(3,4)} (r[s=1]) + \\& \phantom{{}=\frac{1}{(4-2+1)} [T^{OR,2}_{(2,4)} (r[s=1])}T^{OR,2}_{(4,4)} (r[s=1])]\\
& = 17/72.
\end{align*}
Contrast is calculated as  
\begin{align*}
\alpha &= \frac{T^{OR,2}(r[s=0]) - T^{OR,2}(r[s=1])}{1 +T^{OR,2}(r[s=1])} = \frac{\frac{19}{72}-\frac{17}{72}}{1+\frac{17}{72}}
= \frac{2}{89}.
\end{align*}
case 2 : t = 3\\
Average light transmission when $s=0$,
\begin{align*}
T^{OR,3}(r[s=0]) &= \frac{1}{(n-k+1)} [T^{OR,3}_{(k,n)} (r[s=0]) + T^{OR,3}_{(k+1,n)} (r[s=0]) + ..\\&\phantom{{}=\frac{1}{(n-k+1)} [T^{OR,3}_{(k,n)} (r[s=0])}... + T^{OR,3}_{(n,n)} (r[s=0])]\\
&= \frac{1}{(4-2+1)} [T^{OR,3}_{(2,4)} (r[s=0]) + T^{OR,3}_{(3,4)} (r[s=0]) + \\&\phantom{{}= \frac{1}{(4-2+1)} [T^{OR,3}_{(2,4)} (r[s=0]) + T^{OR,3}}T^{OR,3}_{(4,4)} (r[s=0])]\\
&= \frac{1}{3} \times [ \frac {\binom 32}{\binom 42} \times (\frac{1}{2})^{3-1} +  (1-  \frac {\binom 32}{\binom 42})\times (\frac{1}{2})^{3} + \frac {\binom 33}{\binom 43} \times (\frac{1}{2})^{3-1} +\\&\phantom{{}= \frac{1}{3} \times [ \frac {\binom 32}{\binom 42} \times (\frac{1}{2})^{3-1}+(1-  \frac {\binom 32}{\binom 42})}  (1-  \frac {\binom 33}{\binom 43})\times (\frac{1}{2})^{3} + (\frac{1}{2})^{3}]\\
& = 15/96.
\end{align*}
Average light transmission when $s=1$,
\begin{align*}
T^{OR,3}(r[s=1]) &= \frac{1}{(n-k+1)} [T^{OR,3}_{(k,n)} (r[s=1]) + T^{OR,3}_{(k+1,n)} (r[s=1]) + ..\\&\phantom{{}=\frac{1}{(n-k+1)} [T^{OR,3}_{(k,n)} (r[s=1])}... + T^{OR,3}_{(n,n)} (r[s=1])]\\
&= \frac{1}{(4-2+1)} [T^{OR,3}_{(2,4)} (r[s=1]) + T^{OR,3}_{(3,4)} (r[s=1]) + \\&\phantom{{}=\frac{1}{(4-2+1)} [T^{OR,3}_{(2,4)} (r[s=1])}T^{OR,3}_{(4,4)} (r[s=1])]\\
&= \frac{1}{3} \times [(1-  \frac {\binom 32}{\binom 42})\times (\frac{1}{2})^{3} + (1-  \frac {\binom 33}{\binom 43})\times (\frac{1}{2})^{3} + (\frac{1}{2})^{3}]\\
& = 9/96.
\end{align*}
Contrast is calculated as  
\begin{align*}
\alpha &= \frac{T^{OR,3}(r[s=0]) - T^{OR,3}(r[s=1])}{1 +T^{OR,3}(r[s=1])}
= \frac{\frac{15}{96}-\frac{9}{96}}{1+\frac{9}{96}} = \frac{6}{105}. 
\end{align*}
case 3 : t = 4\\
Average light transmission when $s=0$,
\begin{align*}
T^{OR,4}(r[s=0]) &= \frac{1}{(n-k+1)} [T^{OR,4}_{(k,n)} (r[s=0]) + T^{OR,4}_{(k+1,n)} (r[s=0]) + ..\\&\phantom{{}=\frac{1}{(n-k+1)} [T^{OR,4}_{(k,n)} (r[s=0])}... + T^{OR,4}_{(n,n)} (r[s=0])]\\
&= \frac{1}{(4-2+1)} [T^{OR,4}_{(2,4)} (r[s=0]) + T^{OR,4}_{(3,4)} (r[s=0]) +\\&\phantom{{}=\frac{1}{(4-2+1)} [T^{OR,4}_{(2,4)} (r[s=0])} T^{OR,4}_{(4,4)} (r[s=0])]\\
&= \frac{1}{3} \times [ \frac {\binom 42}{\binom 42} \times (\frac{1}{2})^{4-1} +  (1-  \frac {\binom 42}{\binom 42})\times (\frac{1}{2})^{4} + \frac {\binom 43}{\binom 43} \times (\frac{1}{2})^{4-1} +\\&\phantom{{}=\frac{1}{3} \times}  (1-  \frac {\binom 43}{\binom 43})\times (\frac{1}{2})^{4} + \frac {\binom 44}{\binom 44} \times (\frac{1}{2})^{4-1} +  (1-  \frac {\binom 44}{\binom 44})\times (\frac{1}{2})^{4}]\\
& = 1/8.
\end{align*}
Average light transmission when $s=1$,
\begin{align*}
T^{OR,4}(r[s=1]) &= \frac{1}{(n-k+1)} [T^{OR,4}_{(k,n)} (r[s=1]) + T^{OR,4}_{(k+1,n)} (r[s=1]) + ..\\&\phantom{{}=\frac{1}{(n-k+1)} [T^{OR,4}_{(k,n)} (r[s=1])}... + T^{OR,4}_{(n,n)} (r[s=1])]\\
&= \frac{1}{(4-2+1)} [T^{OR,4}_{(2,4)} (r[s=1]) + T^{OR,4}_{(3,4)} (r[s=1]) +\\&\phantom{{}=\frac{1}{(4-2+1)} [T^{OR,4}_{(2,4)} (r[s=1])} T^{OR,4}_{(4,4)} (r[s=1])]\\
&= \frac{1}{3} \times [(1-  \frac {\binom 42}{\binom 42})\times (\frac{1}{2})^{4} + (1-  \frac {\binom 43}{\binom 43})\times (\frac{1}{2})^{4} +\\&\phantom{{}=\frac{1}{3} \times [(1-  \frac {\binom 42}{\binom 42})\times (\frac{1}{2})^{4}} (1-  \frac {\binom 44}{\binom 44})\times (\frac{1}{2})^{4}]\\
&= 0.
\end{align*}
Contrast is calculated as  
\begin{align*}
\alpha &= \frac{T^{OR,4}(r[s=0]) - T^{OR,4}(r[s=1])}{1 +T^{OR,4}(r[s=1])} = \frac{\frac{1}{8}-0}{1+0} = \frac{1}{8}.
\end{align*}
Now, we calculate the value of contrast for (2, 4) VSS scheme by XORed decryption\\
case 1 : t = 2\\  
Average light transmission when $s=0$, 
\begin{align*}
T^{XOR,t}(r[s=0]) &= \frac{1}{(n-k+1)} [T^{XOR,t}_{(k,n)} (r[s=0]) + T^{XOR,t}_{(k+1,n)} (r[s=0]) + ..\\&\phantom{{}=\frac{1}{(n-k+1)} [T^{XOR,t}_{(k,n)} (r[s=0])}... + T^{XOR,t}_{(n,n)} (r[s=0])]\\
&=\frac{1}{(4-2+1)} [T^{XOR,2}_{(2,4)} (r[s=0]) + T^{XOR,2}_{(3,4)} (r[s=0]) +\\&\phantom{{}=\frac{1}{(4-2+1)} [T^{XOR,2}_{(2,4)} (r[s=0])} T^{XOR,2}_{(4,4)} (r[s=0])]\\
&=\frac{1}{(4-2+1)} [ \frac {1}{2} \times (1 + \frac {1}{\binom 42}) +\frac {1}{2}+ \frac {1}{2}]\\
&= 19/36.
\end{align*}
Average light transmission when $s=1$,
\begin{align*}
T^{XOR,t}(r[s=1]) &= \frac{1}{(n-k+1)} [T^{XOR,t}_{(k,n)} (r[s=1]) + T^{XOR,t}_{(k+1,n)} (r[s=1]) + ..\\&\phantom{{}=\frac{1}{(n-k+1)} [T^{XOR,t}_{(k,n)} (r[s=1])}... + T^{XOR,t}_{(n,n)} (r[s=1])]\\
&=\frac{1}{(4-2+1)} [T^{XOR,2}_{(2,4)} (r[s=1]) + T^{XOR,2}_{(3,4)} (r[s=1]) +\\&\phantom{{}=\frac{1}{(4-2+1)} [T^{XOR,2}_{(2,4)} (r[s=1])} T^{XOR,2}_{(4,4)} (r[s=1])]\\
&=\frac{1}{(4-2+1)} [ \frac {1}{2} \times (1 - \frac {1}{\binom 42}) +\frac {1}{2}+ \frac {1}{2}]\\
&= 17/36.
\end{align*}
Contrast is calculated as  
\begin{align*}
\alpha &= \frac{T^{XOR,2}(r[s=0]) - T^{XOR,2}(r[s=1])}{1 +T^{XOR,2}(r[s=1])} = \frac{\frac{19}{36}-\frac{17}{36}}{1+\frac{17}{36}}
= \frac{2}{53}.
\end{align*}
case 2 : t = 3\\
Average light transmission when $s=0$,
\begin{align*}
T^{XOR,t}(r[s=0]) &= \frac{1}{(n-k+1)} [T^{XOR,t}_{(k,n)} (r[s=0]) + T^{XOR,t}_{(k+1,n)} (r[s=0]) + ..\\&\phantom{{}=\frac{1}{(n-k+1)} [T^{XOR,t}_{(k,n)} (r[s=0])}... + T^{XOR,t}_{(n,n)} (r[s=0])]\\
&=\frac{1}{(4-2+1)} [T^{XOR,3}_{(2,4)} (r[s=0]) + T^{XOR,3}_{(3,4)} (r[s=0]) +\\&\phantom{{}=\frac{1}{(4-2+1)} [T^{XOR,3}_{(2,4)} (r[s=0])} T^{XOR,3}_{(4,4)} (r[s=0])]\\
&=\frac{1}{(4-2+1)} [\frac {1}{2}+\frac {1}{2} \times (1 + \frac {1}{\binom 43}) +\frac {1}{2}]\\
&= 13/24.
\end{align*}
Average light transmission when $s=1$,
\begin{align*}
T^{XOR,t}(r[s=1]) &= \frac{1}{(n-k+1)} [T^{XOR,t}_{(k,n)} (r[s=1]) + T^{XOR,t}_{(k+1,n)} (r[s=1]) + ..\\&\phantom{{}=\frac{1}{(n-k+1)} [T^{XOR,t}_{(k,n)} (r[s=1])}... + T^{XOR,t}_{(n,n)} (r[s=1])]\\
&=\frac{1}{(4-2+1)} [T^{XOR,3}_{(2,4)} (r[s=1]) + T^{XOR,3}_{(3,4)} (r[s=1]) + \\&\phantom{{}=\frac{1}{(4-2+1)} [T^{XOR,3}_{(2,4)} (r[s=1])} T^{XOR,3}_{(4,4)} (r[s=1])]\\
&=\frac{1}{(4-2+1)} [\frac {1}{2}+\frac {1}{2} \times (1 - \frac {1}{\binom 43}) +\frac {1}{2}]\\
&= 11/24.
\end{align*}
Contrast is calculated as  
\begin{align*}
\alpha &= \frac{T^{XOR,3}(r[s=0]) - T^{XOR,3}(r[s=1])}{1 +T^{XOR,3}(r[s=1])}=\frac{\frac{13}{24}-\frac{11}{24}}{1+\frac{11}{24}}= \frac{2}{35}.
\end{align*}
case 3 : t = 4\\
Average light transmission when $s=0$,
\begin{align*}
T^{XOR,t}(r[s=0]) &= \frac{1}{(n-k+1)} [T^{XOR,t}_{(k,n)} (r[s=0]) + T^{XOR,t}_{(k+1,n)} (r[s=0]) + ..\\&\phantom{{}=\frac{1}{(n-k+1)} [T^{XOR,t}_{(k,n)} (r[s=0])}... + T^{XOR,t}_{(n,n)} (r[s=0])]\\
&=\frac{1}{(4-2+1)} [T^{XOR,4}_{(2,4)} (r[s=0]) + T^{XOR,4}_{(3,4)} (r[s=0]) +\\&\phantom{{}=\frac{1}{(4-2+1)} [T^{XOR,4}_{(2,4)} (r[s=0])} T^{XOR,4}_{(4,4)} (r[s=0])]\\
&=\frac{1}{(4-2+1)} [\frac {1}{2}+ \frac {1}{2} + \frac {1}{2} \times (1 + \frac {1}{\binom 44})]\\
&= 2/3.
\end{align*}
Average light transmission when $s=1$,
\begin{align*}
T^{XOR,t}(r[s=1]) &= \frac{1}{(n-k+1)} [T^{XOR,t}_{(k,n)} (r[s=1]) + T^{XOR,t}_{(k+1,n)} (r[s=1]) + ..\\&\phantom{{}=\frac{1}{(n-k+1)} [T^{XOR,t}_{(k,n)} (r[s=1])}... + T^{XOR,t}_{(n,n)} (r[s=1])]\\
&=\frac{1}{(4-2+1)} [T^{XOR,4}_{(2,4)} (r[s=1]) + T^{XOR,4}_{(3,4)} (r[s=1]) + \\&\phantom{{}=\frac{1}{(n-k+1)} [T^{XOR,t}_{(k,n)} (r[s=1])} T^{XOR,4}_{(4,4)} (r[s=1])]\\
&=\frac{1}{(4-2+1)} [\frac {1}{2}+ \frac {1}{2} + \frac {1}{2} \times (1 - \frac {1}{\binom 44})]\\
&= 1/3.
\end{align*}
Contrast is calculated as  
\begin{align*}
\alpha &= \frac{T^{XOR,4}(r[s=0]) - T^{XOR,4}(r[s=1])}{1 +T^{XOR,4}(r[s=1])}
= \frac{\frac{2}{3}-\frac{1}{3}}{1+\frac{1}{3}} =  \frac{1}{4}. 
\end{align*}
In \cite{Wu201348}(Table 8) contrast for (3,5) VSS scheme for t = 4 by XOR decryption are depicted incorrectly. So, we calculate the average light transmission when $s=0$
\begin{align*}
T^{XOR,t}(r[s=0]) &= \frac{1}{(n-k+1)} [T^{XOR,t}_{(k,n)} (r[s=0]) + T^{XOR,t}_{(k+1,n)} (r[s=0]) + ..\\&\phantom{{}=\frac{1}{(n-k+1)} [T^{XOR,t}_{(k,n)} (r[s=0])}... + T^{XOR,t}_{(n,n)} (r[s=0])]\\
&=\frac{1}{(5-3+1)} [T^{XOR,4}_{(3,5)} (r[s=0]) + T^{XOR,4}_{(4,5)} (r[s=0])+\\&\phantom{{}=\frac{1}{(5-3+1)} [T^{XOR,4}_{(3,5)} (r[s=0])} T^{XOR,4}_{(5,5)} (r[s=0])]\\
&=\frac{1}{(5-3+1)} [\frac {1}{2} + \frac {1}{2} \times (1 + \frac {1}{\binom 54})+\frac {1}{2} ]\\
&= 8/15.
\end{align*}
Average light transmission when $s=1$,
\begin{align*}
T^{XOR,t}(r[s=1]) &= \frac{1}{(n-k+1)} [T^{XOR,t}_{(k,n)} (r[s=1]) + T^{XOR,t}_{(k+1,n)} (r[s=1]) +  ..\\&\phantom{{}=\frac{1}{(n-k+1)} [T^{XOR,t}_{(k,n)} (r[s=1])}... + T^{XOR,t}_{(n,n)} (r[s=1])]\\
&=\frac{1}{(5-3+1)} [T^{XOR,4}_{(3,5)} (r[s=0]) + T^{XOR,4}_{(4,5)} (r[s=0])+\\&\phantom{{}=\frac{1}{(5-3+1)} [T^{XOR,4}_{(3,5)} (r[s=0])} T^{XOR,4}_{(5,5)} (r[s=0])]\\
&=\frac{1}{(5-3+1)} [\frac {1}{2} + \frac {1}{2} \times (1 - \frac {1}{\binom 54})+\frac {1}{2} ]\\
&= 7/15.
\end{align*}
Contrast is calculated as  
\begin{align*}
\alpha &= \frac{T^{XOR,4}(r[s=0]) - T^{XOR,4}(r[s=1])}{1 +T^{XOR,4}(r[s=1])} = \frac{\frac{8}{15}-\frac{7}{15}}{1+\frac{7}{15}} = \frac{1}{22}. 
\end{align*}
In \cite{Wu201348}(Table 9) contrast for (4, 5) VSS scheme for t = 4 by XOR decryption are depicted incorrectly. So, we calculate the average light transmission when $s=0$
\begin{align*}
T^{XOR,t}(r[s=0]) &= \frac{1}{(n-k+1)} [T^{XOR,t}_{(k,n)} (r[s=0]) + T^{XOR,t}_{(k+1,n)} (r[s=0]) +  ..\\&\phantom{{}=\frac{1}{(n-k+1)} [T^{XOR,t}_{(k,n)} (r[s=0])}... + T^{XOR,t}_{(n,n)} (r[s=0])]\\
&=\frac{1}{(5-4+1)} [T^{XOR,4}_{(4,5)} (r[s=0]) +  T^{XOR,4}_{(5,5)} (r[s=0])]\\
&=\frac{1}{(5-4+1)} [ \frac {1}{2} \times (1 + \frac {1}{\binom 54})+\frac {1}{2} ]\\
&= 11/20.
\end{align*}
Average light transmission when $s=1$,
\begin{align*}
T^{XOR,t}(r[s=1]) &= \frac{1}{(n-k+1)} [T^{XOR,t}_{(k,n)} (r[s=1]) + T^{XOR,t}_{(k+1,n)} (r[s=1]) + ..\\&\phantom{{}=\frac{1}{(n-k+1)} [T^{XOR,t}_{(k,n)} (r[s=1])}... + T^{XOR,t}_{(n,n)} (r[s=1])]\\
&=\frac{1}{(5-4+1)} [T^{XOR,4}_{(4,5)} (r[s=1]) +  T^{XOR,4}_{(5,5)} (r[s=1])]\\
&=\frac{1}{(5-4+1)} [ \frac {1}{2} \times (1 - \frac {1}{\binom 54})+\frac {1}{2} ]\\
&= 9/20.
\end{align*}
Contrast is calculated as  
\begin{align*}
\alpha &= \frac{T^{XOR,4}(r[s=0]) - T^{XOR,4}(r[s=1])}{1 +T^{XOR,4}(r[s=1])} =\frac{\frac{11}{20}-\frac{9}{20}}{1+\frac{9}{20}} = \frac{2}{29}. 
\end{align*}

\section*{References}
\bibliographystyle{plain}
\bibliography{reference}

\begin{thebibliography}{2}
\providecommand{\natexlab}[1]{#1}
\providecommand{\url}[1]{\texttt{#1}}
\expandafter\ifx\csname urlstyle\endcsname\relax
  \providecommand{\doi}[1]{doi: #1}\else
  \providecommand{\doi}{doi: \begingroup \urlstyle{rm}\Url}\fi

\bibitem[Chen and Tsao(2011)]{Chen20111197}
Tzung-Her Chen and Kai-Hsiang Tsao.
\newblock Threshold visual secret sharing by random grids.
\newblock \emph{Journal of Systems and Software}, 84\penalty0 (7):\penalty0
  1197 -- 1208, 2011.
\newblock ISSN 0164-1212.
\newblock \doi{http://dx.doi.org/10.1016/j.jss.2011.02.023}.
\newblock URL
  \url{http://www.sciencedirect.com/science/article/pii/S0164121211000513}.

\bibitem[Wu and Sun(2013)]{Wu201348}
Xiaotian Wu and Wei Sun.
\newblock Random grid-based visual secret sharing with abilities of or and xor
  decryptions.
\newblock \emph{Journal of Visual Communication and Image Representation},
  24\penalty0 (1):\penalty0 48 -- 62, 2013.
\newblock ISSN 1047-3203.
\newblock \doi{http://dx.doi.org/10.1016/j.jvcir.2012.11.001}.
\newblock URL
  \url{http://www.sciencedirect.com/science/article/pii/S1047320312001642}.

\end{thebibliography}

\end{document}